\documentclass[3p,number,a4paper,sort&compress]{elsarticle}
\usepackage{graphicx}
\usepackage{epstopdf}
\usepackage{dcolumn}
\usepackage{amsmath}
\usepackage{mathptmx, courier, pifont}
\usepackage[scaled=0.92]{helvet}
\usepackage[T1]{fontenc}
\usepackage{textcomp}
\usepackage{color}
\usepackage{lineno,hyperref}
\modulolinenumbers[1]

\begin{document}
\begin{frontmatter}

\title{Accurate correction of magnetic field instabilities for high-resolution isochronous mass measurements in storage rings}

\author[USTC,IMP]{P.~Shuai}
\cortext[mycorrespondingauthor]{Corresponding authors}
\author[IMP]{H.~S.~Xu\corref{mycorrespondingauthor}}
\ead{hushan@impcas.ac.cn}
\author[IMP]{Y.~H.~Zhang\corref{mycorrespondingauthor}}
\ead{yhzhang@impcas.ac.cn}
\author[IMP,GSI,Max-Planck]{Yu.~A. Litvinov\corref{mycorrespondingauthor}}
\ead{Y.Litvinov@GSI.de}
\author[IMP,ORSAY]{M.~Wang}
\author[IMP,GSI,Max-Planck]{X.~L.~Tu}
\author[Max-Planck]{K.~Blaum}
\author[IMP]{X.~H.~Zhou}
\author[IMP]{Y.~J.~Yuan}
\author[ORSAY]{G.~Audi}
\author[IMP,GUCAS,Max-Planck]{X.~L.~Yan}
\author[IMP,GUCAS]{X.~C.~Chen}
\author[IMP,GUCAS]{X.~Xu}
\author[IMP,GUCAS]{W.~Zhang}
\author[Beihang]{B.~H.~Sun}
\author[SAITAMA]{T.~Yamaguchi}

\author[IMP]{R.~J.~Chen}
\author[IMP,GUCAS]{C.~Y.~Fu}
\author[IMP,GUCAS]{Z.~Ge}
\author[IMP,GUCAS]{W.~J.~Huang}
\author[IMP,GUCAS]{D.~W.~Liu}
\author[IMP,GUCAS]{Y.~M.~Xing}
\author[USTC,IMP]{Q.~Zeng}

\address[USTC]{Research Center for Hadron Physics, National Laboratory of Heavy Ion Accelerator Facility in Lanzhou and University of Science and Technology of China, Hefei 230026, People's Republic of China}
\address[IMP]{Key Laboratory of High Precision Nuclear Spectroscopy, Center for Nuclear Matter Science, Institute of Modern Physics, Chinese Academy of Sciences, Lanzhou 730000, People's Republic of China}
\address[GSI]{GSI Helmholtzzentrum f\"{u}r Schwerionenforschung, Planckstra{\ss}e 1, 64291 Darmstadt, Germany}
\address[Max-Planck]{Max-Planck-Institut f\"{u}r Kernphysik, Saupfercheckweg 1, 69117 Heidelberg, Germany}
\address[ORSAY]{CSNSM-IN2P3-CNRS, Universit\'{e} de Paris Sud, F-91405 Orsay, France}
\address[GUCAS]{Graduate University of Chinese Academy of Sciences, Beijing, 100049, People's Republic of China}
\address[Beihang]{School of Physics and Nuclear Engineering, Beihang University, Beijing 100191, People's Republic of China}
\address[SAITAMA]{Department of Physics, Saitama University, Saitama 338-8570, Japan}

\begin{abstract}
Isochronous mass spectrometry (IMS) in storage rings is a successful technique
for accurate mass measurements of short-lived nuclides with relative precision of about $10^{-5}-10^{-7}$.
Instabilities of the magnetic fields in storage rings are one of the major contributions limiting the achievable mass resolving power,
which is directly related to the precision of the obtained mass values.
A new data analysis method is proposed allowing one to minimise the effect of such instabilities.
The masses of the previously measured at the CSRe $^{41}$Ti, $^{43}$V, 
$^{47}$Mn, $^{49}$Fe, $^{53}$Ni and $^{55}$Cu nuclides were re-determined with this method.
An improvement of the mass precision by a factor of $\sim 1.7$ has been achieved for $^{41}$Ti and $^{43}$V.
The method can be applied to any isochronous mass experiment irrespective of the accelerator facility.
Furthermore, the method can be used as an on-line tool for checking the isochronous conditions of the storage ring.
\end{abstract}

\begin{keyword}
\texttt ~Isochronous mass spectrometry\sep Ion storage rings\sep Magnetic field instabilities
\end{keyword}

\end{frontmatter}



\section{Introduction}
Nuclear masses are fundamental properties of atomic nuclei, the knowledge of which is indispensable for nuclear structure and astrophysics research~\cite{Blaum1,Lunney,Grawe07}.
Therefore, precise mass measurements are highly important and are pursued by many groups worldwide~\cite{100yms}.
Presently, nuclides with unknown masses lie far-off stability and their mass measurements become a technical challenge due to the tiny production rates and short half-lives.
The isochronous mass spectrometry (IMS), applied to relativistic nuclear reaction products, is an ideal technique for mass measurements of such short-lived rare nuclides~\cite{EMIS,BLT}.
The IMS measurements were pioneered at the ESR facility of GSI~\cite{FGM,Sun1,Hausmann1,Stadlmann1} and
have recently been established at the storage ring CSRe of IMP~\cite{Tu11,Xu},

On the one hand, the IMS has an ultimate sensitivity to single ions stored in a storage ring.
On the other hand, the IMS does not require any lengthy preparation of the particles,
like, e.g., cooling, and can be applied on systems with lifetimes as short as a few tens of microseconds~\cite{SunPLB}.

We note, that the secondary ions of interest have an inevitable velocity spread, $\Delta{v}/v$, due to the production reaction process.
The revolution period, $T$, of an ion circulating in a storage ring depends on its mass-to-charge ratio, $m/q$, and on its velocity, $v$.
In a certain magnetic rigidity acceptance, the ions with different $m/q$ and $v$ can be stored,
where, for the same type of ions, the faster ions circulate on longer orbits while the slower ones on the shorter orbits.
Just in a special, isochronous ion-optical setting of the ring, the differences of the orbit lengths compensate the differences
of velocities, although within a limited range of magnetic rigidities, $\Delta (B\rho)/B\rho$.
Then, the revolution periods of stored ions depend only on their $m/q$ ratios and are independent of their velocities.
Thus, the masses of stored ions can be measured through accurate determination of their revolution period $T$.

Obviously, the precision of the revolution period determinations is crucial.
For a certain ion species with the same mass-to-charge ratio ($m/q=const$), the resolving power $\delta T/T$ is given in first-order approximation by the frequency dispersion
($\eta$) and the relative spread of magnetic rigidities, $\delta (B\rho)/B\rho$, according to the following equation~\cite{Haus00}:
\begin{equation}
\frac{\delta{T}}{T}
=-\eta \frac{\delta(B\rho)}{B\rho}
=-\biggl(1-\frac{\gamma^2}{\gamma_{t}^2}\biggr)\frac{\delta v}{v}, \label{eq1}
\end{equation}
where $\gamma$ is the relativistic Lorentz factor and $\gamma_t$ is the transition energy of a storage ring,
which connects the relative change of the orbit length to the relative change of magnetic rigidity of the circulating ions.
In a recent experiment in the CSRe, in order to achieve the resolving power $\delta T/T \sim 10^{-6}$,
$\eta$ and $\delta (B\rho)/B\rho$ were restricted to be within $\sim 10^{-3}$.
From Eq.~(\ref{eq1}) one sees, that $\delta T$ depends on the absolute value of $T$ (or, in other words, on the $m/q$ value) and varies with the ion's velocities.
In the CSRe experiments, $\delta T $ typically reaches its minimum at $\sim 2$ ps, which is called good isochronous region corresponding to $\gamma \sim \gamma_t$~\cite{Xu}.
Individual measurements of the revolution periods of the stored ions are extremely fast and are $\sim1-2$~ms and $200~\mu$s in the GSI and IMP measurements, respectively,
which allows for addressing nuclides with half-lives at least as short as the required measurement time.

In discussions above, the magnetic fields of the storage ring magnets are assumed perfectly constant.
However, in reality, the fields of the CSRe magnets experience slow fluctuations of the order of $\delta B/B\sim10^{-5}-10^{-4}$.
As a consequence, the revolution periods of the ions stored in the ring inevitably change too, leading to an extra time spread of more than $10$~ps.
This huge time spread significantly deteriorates the achievable resolving power.
Although the magnetic fields can be stabilised down to $\delta B/B \sim 10^{-6}$, as, e.g., in RIKEN~\cite{Yamaguchi},
by using dedicated power supply systems, the instabilities of the fields are in principle unavoidable.
Therefore, much effort was devoted to find a proper data analysis method to eliminate the influence of the magnetic field instabilities~\cite{Tu11,Zhang12}.

Revolution periods change due to magnetic field variations by approximately the same value for all ions,
that is, the change in the magnetic field causes a drift of the overall spectrum, which can be corrected for.
Since the pioneering experiments at GSI~\cite{Haus00,Hausmann1,Stadlmann1,Sun1}, the methods for correcting the magnetic drift are being developed all the time.
In the first IMS experiments, the data acquired during a short period of time (typically several minutes) were grouped together in a sub-spectrum.
The accumulated statistics allowed for a least squares type analysis of the drifts between individual sub-spectra, which allowed for merging them together into a final spectrum.
Afterwards, the mean revolution times and standard deviations of the peaks in the total spectrum are used to obtain the mass values and their uncertainties~\cite{Hausmann1,Zhang12}.
The disadvantage of such method is that the magnetic field instabilities within the accumulation time are not taken into account.
Another method is the correlation-matrix approach, which was developed at first for the Schottky mass measurements at GSI~\cite{Radon} and extended later also to the IMS~\cite{Sun1}.
It was applied only once on uranium fission data, where, due to the tiny secondary particle yields, the combination of several spectra together was still needed.

To overcome the disadvantage of grouping the data, the data on the mass measurements of $^{78}$Kr fragments performed at CSRe~\cite{Tu11,Xu} were analysed in another way:
The drifts due to magnetic field instabilities were corrected between individual measurements, where the instabilities within the measurement time ($200~\mu$s) can safely be neglected.
Seven ions, which have the highest counting statistics, were chosen as references.
The revolution periods for the reference ions were set to the same value in all measurements where these ions were present, which combined together form a sub-spectrum for this reference ion.
The disadvantage of this method is clearly the increase of the revolution period spreads due to the artificial setting of the spread for the reference ion to zero.
Furthermore, the measurements not containing reference ions were not considered in the analysis.
Nevertheless, a much higher resolving power could be achieved with this method as compared to an uncorrected spectrum.
However, this method may not be suitable for experiments in which no suitable reference ions are expected, e.g., for mass measurements of very neutron-rich nuclei.

In this work, a new method, taking into account the drifts between individual measurements, is proposed to accurately deduce the revolution periods of the stored ions and their standard deviations.

\section{Data analysis method}
A typical IMS experiment lasts for several days or weeks.
The revolution periods of several tens of ions can be determined in several ten thousands of measurements.
Each individual measurement contains around ten simultaneously stored ions and takes merely 200 $\mu$s.
Within such short period of time, the magnetic fields can be regarded as being constant.
This is an important starting point in our analysis and discussions.
The experimental details, ion identification, and extraction of the revolution periods from acquired timing information on the circulating ions can be found in Refs.~\cite{Tu11,Zhang12}.


Let us define the revolution periods of the $i$-th ion species ($i=1,2,3,\cdots,N_{s}$, with $N_s$ being the total number of different ion species)
in the $j$-th measurement ($j=1,2,3,\cdots,N_{m}$, with $N_m$ being the total number of measurements) as $T_{i,j}$.
The magnetic fields of the storage ring vary slowly in time around a central mean value $B_0$.
The variation of the magnetic field in the $j$-th measurement with respect to $B_0$ is $\delta B_j=B_j-B_0$.
The $T_{i,j}$ data in this measurement can be written as:
\begin{equation}
T_{i,j}=T_{i,j}^{0} - \Delta T_j,
\label{eq2}
\end{equation}
where $T_{i,j}^{0}$ are the revolution periods in the absence of magnetic field variations and $\Delta T_j$ is the drift due to $\delta B_j$. 

The $T^0_{i,j}$ follow the normal distribution, $N(\mu_{T_{i}},\sigma_{T_{i}}^{2})$ with expectation mean values $\mu_{T_{i}}$ and their standard deviations $\sigma_{T_{i}}$. 
Here, the $\sigma_{T_{i}}$ include all the other uncertainty factors of the measurements except for the instabilities of magnetic fields. 
In real measurements, due to the isochronicity conditions, the $\sigma_{T_{i}}$ depend on $\mu_{T_{i}}$.
The goal of the following analysis procedure is to determine the $\mu_{T_{i}}$ and $\sigma_{T_{i}}$ for all ion species $(i=1,2,3,\cdots,N_{s})$, which can then be used for the mass determination.
Details of the mathematical derivations can be found in Ref.~\cite{ShuaiPHD}.

Here it is worth noting, that $T_{i,j}^0$ and  $\Delta T_j$ are independent variables and they vary randomly in the long-time measurements around $\mu_{T_{i}}$ and 0, respectively. 
In an ideal case when $\Delta B_j=0$ ($j=1,2,3,\cdots,N_{m}$), all $\Delta T_j=0$ and the measured mean revolution periods $T_i=\langle T_{i,j}\rangle$ should approach $\mu_{T_{i}}$
with experimental standard deviations $s_{T_i}$ approaching $\sigma_{T_{i}}$.
For several ten thousands of measurements the fluctuations of the magnetic fields around $B_0$ average out and one can write:
\begin{equation}
T_i=\langle T_{i,j}\rangle=\langle T_{i,j}^0\rangle \rightarrow\mu_{T_i},
\label{eq3a}
\end{equation}
However, the standard deviations include the contribution due to magnetic field fluctuations, $\sigma_{B}$:
\begin{equation}
s_{T_i}^2\rightarrow\sigma_{T_i}^2+\sigma_{B}^2.
\label{eq3b}
\end{equation}

Let us assume that all the $N_s$ ions are present in each measurement $j$.
If the $\mu_{T_{i}}$ and $\sigma_{T_{i}}$ are known, then
we can re-construct a new variable $T_{i,j}^{\prime}$, which can be calculated directly from experimental data, through a linear transformation of the variable $T_{i,j}$, i.e.,

\begin{equation}
\begin{split}
&	T_{i,j}^{\prime}
=T_{i,j}-\sum_{i=1}^{N_s} \frac{1}{\sigma_{T_i}^{2}}(T_{i,j}-\mu_{T_i}) / \sum_{i=1}^{N_s}
\frac{1}{\sigma_{T_i}^{2}}	\\
& \qquad\qquad(i=1,2,3, \cdots, N_s,~~~ j=1,2,3, \cdots,N_m).\label{eq5}
\end{split}
\end{equation}
In case of the $N_s \rightarrow \infty$, the standard deviations of $T_{i}^{\prime}$ distributions should approach $\sigma_{T_i}$.
However, since the number of ion species, $N_s$, is finite, the standard deviations $s_{T_i^\prime}$ of the resulting distributions $T_{i}^{\prime}$ are smaller than $\sigma_{T_{i}}$.
In other words, the distributions $T_{i}^{\prime}$ are ``overcorrected''. For more details see Ref.~\cite{ShuaiPHD}.

Using Eq.~(\ref{eq2}) and given the fact that $\Delta T_j$ is a constant for all the ion species in the $j-$th measurement, it can be shown that $T_{i,j}^{\prime}$ are related to $T_{i,j}^0$ as:
\begin{equation}
\begin{split}
&	T_{i,j}^{\prime}
=T_{i,j}^0-\sum_{i=1}^{N_s} \frac{1}{\sigma_{T_i}^{2}}(T_{i,j}^{0}-\mu_{T_i}) / \sum_{i=1}^{N_s} \frac{1}{\sigma_{T_i}^{2}}
=T_{i,j}^0-\delta T_j \\
& \qquad\qquad (i=1,2,3, \cdots, N_s,~~~ j=1,2,3, \cdots, N_m). \label{eq6}
\end{split}
\end{equation}
where $\delta T_j$ is the term accounting for the ``overcorrection'' of the $T_{i}^{\prime}$ distributions.
The variance of the $\delta T_j$ values is given by:
\begin{equation}
V(\delta T_j)=1/\sum_{i=1}^{N_s} \frac{1}{\sigma_{T_{i}}^{2}}, 
\end{equation}
which approaches 0 when $N_s \rightarrow \infty$.

%

Under the condition that $T_{i,j}^\prime$ and $\delta T_j$ are independent variables one can write:
\begin{equation}
T_i^{\prime}=\langle T_{i,j}^{\prime}\rangle=T_i=\langle T_{i,j}\rangle=\langle T_{i,j}^{0}\rangle=\mu_{T_i} ~~~~~ (i=1,2,3, \cdots, N_s), \label{eq8}
\end{equation}
and 
\begin{equation}
\sigma_{T_{i}}^{2} = s_{T_{i}^{\prime}}^{2} + V(\delta T_j) ~~~~~~~~~~~~ (i=1,2,3, \cdots, N_s). \label{eq9}
\end{equation}

As seen in Eq.~(\ref{eq9}), $s_{T_{i}^{\prime}}\neq \sigma_{T_{i}}$, which may cause extra systematic errors:
In an ideal case when all $\delta B_j=0$ and assuming all $\sigma_{T_{i}}=\sigma_T=const$, 
all $T_{i}^{\prime} =T_{i}$ but $\sigma_{T}^{2} = s_{T_{i}^{\prime}}^{2} + \sigma_T^2/N_s$.
This means that the uncertainties are overestimated by $\sigma_T^2/N_s$.

For realistic magnitudes of $\delta B_j$ in the CSRe experiments, 
$V(\delta T_j) << \sigma_{B}^2$, and the new method results in a significant improvement of the resolving power.

To find the unknown $\mu_{T_{i}}$ and $\sigma_{T_{i}}$ in a real data analysis, the Eqs.~(\ref{eq5})-(\ref{eq9}) are solved iteratively.
The important starting step is to provide externally an initial reference set of $\mu^*_{T_{i}}$ and $\sigma^*_{T_{i}}$.
For this purpose, simulated values of the revolution periods~\cite{Hausmann1,Yuri05,Tu11} or the ones obtained from the raw experimental data (see Eqs.~(\ref{eq3a}) and (\ref{eq3b})) can be used.

The second step is to calculate the $T_{i,j}^{\prime}$ values for all ions present in all individual measurements.
This can be done with Eq.~(\ref{eq5}) 
by substituting $\mu_{T_{i}}$ and $\sigma_{T_{i}}$ in Eq.~(\ref{eq5}) with $\mu^*_{T_{i}}$ and $\sigma^*_{T_{i}}$. 
From the set of $T_{i,j}^{\prime}$ values, one can calculate the mean revolution periods, $T_i^{\prime}$, and the corresponding standard deviations, $s_{T_i^{\prime}}$ \cite{ShuaiPHD}:
\begin{equation}
\begin{split}
&	T_{i}^{\prime}=\mu_{T_i}^{\prime} =\frac{\sum_{j} T_{i,j}^{\prime}}{M_{i}},~~~
s_{T_i^{\prime}}^{2}=\frac{\sum_{j} (T_{i,j}^{\prime}-\mu_{T_i}^{\prime})^{2}}{M_{i}-1}	\\
& \qquad\qquad (i=1,2,3, \cdots, N_s), \label{eq10}
\end{split}
\end{equation}
where $M_{i}$ is the number of occurrences of the $i$-th ion species in all $N_m$ measurements.
We note, that the calculated $s_{T_i^{\prime}}^{2}$ values with Eq.~(\ref{eq10}) are different from those calculated with Eq.~(\ref{eq9}).
They would be the same if the initial parameters $\mu^*_{T_i}$ and $\sigma^*_{T_i}$ are exactly equal to $\mu_{T_{i}}$ and $\sigma_{T_{i}}$, respectively.

Therefore, in the next step, the results of Eq.~(\ref{eq10}) are used as the new $\mu^*_{T_i} = T_{i}^{\prime}$ and $\sigma^{*2}_{T_{i}} = s^2_{T_i^{\prime}}+V(\delta T_j)$.
This procedure is performed until the convergence is reached.
Thus the last $\mu^*_{T_i}$ and $\sigma^{*2}_{T_{i}} $ values
correspond to the best approximation, as allowed by the statistics in a particular experiment, of the expected $\mu_{T_i}$ and $\sigma_{T_{i}} $ which can be used for the mass determination.




In reality the statistics is finite and the number of species in each measurement is not equal to $N_s$.
For example, three different ions of species $A,~B,~C$ are present in the first measurement while the species $C,~D,~E,~F$ are stored in a subsequent measurement.
Furthermore all different ions appear randomly in different measurements, and finally they have different statistics dependent, e.g., on the production cross-section, transmission efficiency, etc.
Therefore, $V(\delta T_j)=1/ \sum_{i=1}^k 1/\sigma_{T_{i}}^{2}$ can significantly be different for measurements with significantly different number of present ion species $k$.
The number of individual measurements $N_m$ is several ten thousands while the average number of simultaneously stored ion species $k$ is about 10.
One can show mathematically (see Ref.~\cite{ShuaiPHD}) that in the limit of $M_i\rightarrow\infty$,
$(\sigma_{T_i}^{*})^{2}=s_{T_i^{\prime}}^{2}+V(\delta T_{j})$
and
$(\sigma_{T_i}^{*})^{2}=s_{T_i^{\prime}}^{2}+V_{i}(\delta T_{j})$
provide the same result, where $V_{i}(\delta T_{j})$ is a value for the $i$-th ion species obtained by averaging of $V(\delta T_j)$ over all measurements where the $i$-th ion species is present ($i=true$):
\begin{equation}
V_i(\delta T_j)=\frac{\sum_{j} V(\delta T_{j})|_{i=true}}{M_{i}}~~~~~~~~~~(i=1,2,3,\cdots,N_s), \label{eq11}
\end{equation}
Thus the $\sigma^{*2}_{T_{i}} = s^2_{T_i^{\prime}}+V_i(\delta T_j)$ were used in the iteration procedure above.

\section{Application of the method to real experimental data}

The data analysis method was applied to the IMS experimental data obtained at the CSRe and results of which were published in Ref.~\cite{Yan13}.
By using the new method, the re-determined mass values are consistent with the previously published results, and for some nuclei at isochronous region 
the mass precision is higher than previous one.
Furthermore, since the initial $\mu^*_{T_{i}}$ and $\sigma^*_{T_{i}}$ sets can be provided from an external simulated spectrum, the new method can be an excellent on-line
tool to provide a quick response on the experimental conditions such as, e.g., whether the $B\rho$ setting of the ring is tuned correctly to achieve the "good" isochronous conditions for the nuclides of interest.
To illustrate the feasibility and reliability of the method, we present the newly-obtained results from the experiment aiming at mass measurements of $^{58}$Ni projectile fragments.
The experimental details are reported in Refs.~\cite{Zhang12,Yan13}.

In this experiment, a total number $N_m=15 549$ of measurements including 195 869 ions of $N_s=131$ ion species were acquired.
Magnetic fields were constantly monitored through Hall-probe measurements.
In the original data analysis~\cite{Zhang12,Yan13}, the regions of the nearly constant magnetic fields were identified and the corresponding measurements were combined together in sub-spectra.
The sub-spectra were then merged together by finding the best possible overlap between them forming the final revolution time spectrum.
The particle identification was realized by comparing this final spectrum with a simulated one.

In our analysis, we considered all unambiguously identified ions.
For the initial $\mu^*_{T_{i}}$ and $\sigma^*_{T_{i}}$ sets we employed the simulated $T_i$ and $\sigma _i$ for each ion species.
It is clear that in the first few iterations the $\mu_{T_i}^{*}$ and $\sigma _{T_i}^{*}$ $(i=1,2,3,\cdots,N_s)$ values deviate significantly from the initial ones.
The convergence is reached within a few tens of iterations.

\begin{figure}\centering
\includegraphics[angle=0,width=8.5 cm]{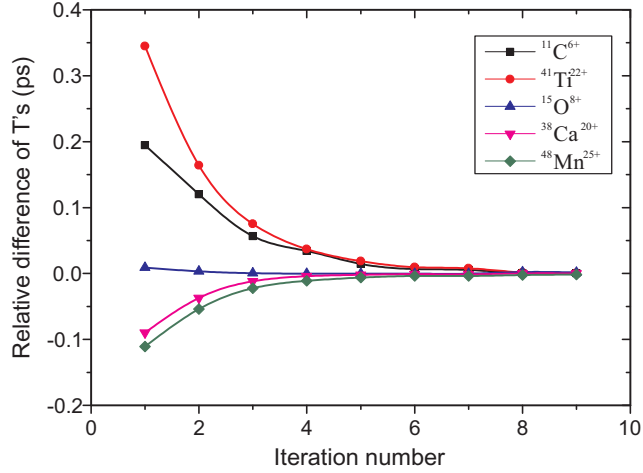}
\caption{(Colour online) The convergence of $\mu_{T_i}^{\prime}$ values for several example ions versus the iteration number. 
\label{Fig01}}
\end{figure}

Figure~\ref{Fig01} presents the convergence of $\mu_{T_i}^{*}$ on the example of several ionic species with different mean $T_i$.
The latter were chosen to illustrate the convergence in the "good" isochronous region as well as far away from it.
We observe nearly exponential convergence of the $\mu_{T_i}^{*}$ values versus the iteration number.

\begin{figure}\centering
\includegraphics[angle=0,width=8.5 cm]{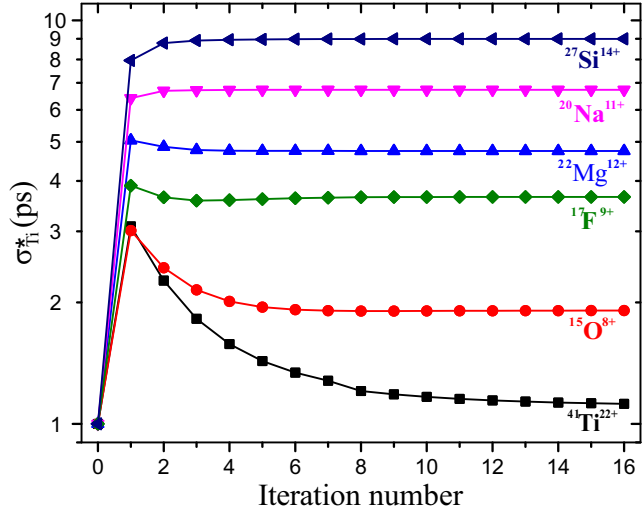}
\caption{(Colour online) The convergence of the standard deviations $\sigma _{T_i}^{*}$ for several example nuclides versus the iteration number. The initial values of the standard deviations were set equal for all ions. 
\label{Fig02}}
\end{figure}

Figure~\ref{Fig02} shows the convergence of the $\sigma _{T_i}^{*}$ values.
In contrast to $\mu_{T_i}^{*}$ values, the convergence of $\sigma _{T_i}^{*}$ values is rather slow.
Typically, a few tens of iterations are sufficient to achieve the variation between subsequent iterations of less than $10^{-3}$~ps for both
$\mu_{T_i}^{*}$ and $\sigma _{T_i}^{*}$ $(i=1,2,3,\cdots,~N_s)$.
The final values are considered to be the ``true'' $\mu_{T_i}$ and $\sigma _{T_i}$.

\begin{figure}\centering
\includegraphics[angle=0,width=8.5 cm]{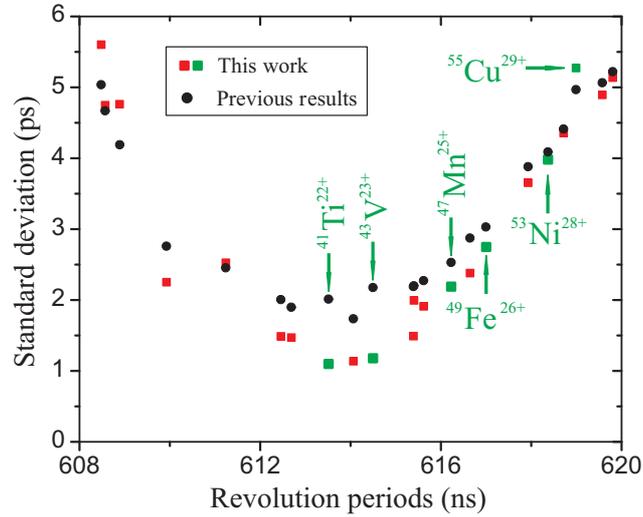}
\caption{(Colour online) The comparison of the standard deviations obtained in this work (red or green filled squares) and in Refs.~\cite{Zhang12,Yan13} (black filled circles).
A range of revolution periods of $608~{\rm ns} \le t \le 620$~ns is considered.
Nucleus with masses unknown according to the AME'11 evaluation~\cite{AME2011} are labelled with the corresponding identification.
\label{Fig03}}
\end{figure}

Figure~\ref{Fig03} provides a comparison of the obtained ``true'' $\sigma _{T_i}$ values to those taken from Refs.~\cite{Zhang12,Yan13} for the same ion species.
As expected, in the ``good'' isochronous region, the new $\sigma _{T_i}$ values are smaller than those from Refs.~\cite{Zhang12,Yan13}.
This result is easy to understand since the additional uncertainty $\sigma_B$ is to a large extend (or even completely) removed in our new data analysis.
Obviously this leads to a higher precision mass determination in this region.

The original aim in the experiment was to set the ``good'' isochronous region on the $^{47}$Mn ions.
However, Figure~\ref{Fig03} shows that the best isochronous
condition ($|\gamma - \gamma_t|=min$) are fulfilled for $^{41}$Ti ions, and that the $\sigma _{T}$ for $^{47}$Mn ions is about twice of that for $^{41}$Ti ions.
This mismatch translates into a significant increase of statistics, and accordingly the beam time duration, which has to be accumulated for $^{47}$Mn ions to achieve the aimed mass uncertainties.
This proves that the present method can be valuable as the on-line monitoring tool for checking the correctness of the isochronous setting of the ring.
By going to more exotic nuclides with lower production yields, the on-line control of the ring settings will become more and more important.
We emphasise that the method can be applied to any IMS data irrespective of the storage ring facility where the data are acquired.

\begin{figure}\centering
 \includegraphics[angle=0,width=8.5 cm]{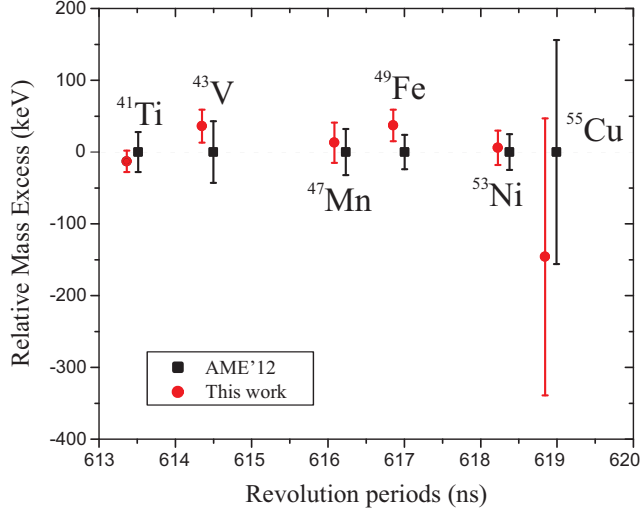}
\caption{(Colour online) Comparison of the masses obtained in Refs.~\cite{Zhang12,Yan13} (black filled circles) and in this work (red filed squares). As a reference, the mass excess values from the AME'12 evaluation~\cite{AME12} are used.
Note that, the precision of the obtained mass values is improved for all considered nuclides except for $^{55}$Cu. The latter is mainly due to very scarce counting statistics for this nuclide.
\label{Fig04}}
\end{figure}

Finally, we used the new $\mu_{T_i}^{*}$ and $\sigma _{T_i}^{*}$ values deduced from the present analysis to re-determine the masses of interest.
We used the same reference ions for the calibration as well as the fitting procedure as in Refs.~\cite{Zhang12,Yan13}.
The re-determined mass excess values and their uncertainties are given in Table.~\ref{table01}.
A comparison of the re-determined values with the ones from Refs.~\cite{Zhang12,Yan13} is illustrated in Figure~\ref{Fig04}.
All data points agree within $1\sigma$ confidence level.
However, a significant improvement of the precision for the masses of $^{41}$Ti and $^{43}$V is achieved.
The mass resolving power of $^{41}$Ti is calculated to be $R=m/\Delta m \approx 310 000$ (sigma) which is increased by a factor of $\sim1.7$ compared to the previously published one.

\begin{table}\centering
\caption{The mass values for $T_{z}=-3/2$ nuclides obtained in this work. The columns list from left to right: the nuclide, the number of identified ions $M_i$, $ME$ values from Refs.~\cite{Zhang12,Yan13}, $ME$ values obtained in this work,
and the difference of these mass excess results $\Delta ME$.}
\begin{tabular}{ccccc}
\hline\hline
$Nuclide$ &  $Counts$  & $ME$ \cite{Zhang12,Yan13}  & $ME$ (this work)      & $\Delta ME$ \\
		  &            & (keV)                      & (keV)     			& (keV)  \\[1mm]
\hline
$^{41}$Ti    &  76     & $-15698(28)$      			& $-15711(15)$      	& $-13(28)$ \\
$^{43}$V     &  42     & $-17916(43)$      			& $-17880(23)$      	& $36(43)$\\
$^{47}$Mn    &  119    & $-22566(32)$      			& $-22553(28)$      	& $13(32)$ \\
$^{49}$Fe    &  335    & $-24751(24)$      			& $-24714(22)$      	& $37(24)$ \\
$^{53}$Ni    &  647    & $-29631(25)$      			& $-29625(24)$      	& $6(25)$ \\
$^{55}$Cu    &  19     & $-31635(156)$     			& $-31781(193)$     	& $-146(193)$ \\
\hline\hline
\end{tabular}
{\normalsize }
\label{table01}
\end{table}

\section{Summary}

The magnetic field instabilities cause a serious deterioration of the mass resolving power in isochronous mass measurements in storage rings, which in turn reduces the achievable precision of the measured mass values.
The instabilities are slow and seen as an extra broadening of the revolution period peaks in the measured spectra.
However, the magnetic fields of the ring can be regarded as being constant during a short time required to perform individual measurements of revolution periods of stored ions.
This time is merely $\sim 200~\mu$s.
The measurements are associated with individual injections of new ions into the ring which are done every few seconds.
The instabilities of magnetic fields cause a shift of the entire revolution period spectrum between such individual measurements.
With a new method, the overall data obtained in the experiment are used to determine these shifts and thus cancel the influence of the magnetic field instabilities.

The mean values and the standard deviations, including the contribution due to unstable magnetic fields, of the measured revolution periods are connected via a set of equations.
In our method this set of equations is solved iteratively providing the mean revolution periods and the standard deviations without the latter contribution.
These values are then used for the mass determination in a standard way.

The new method has been applied to previously published data from an experiment performed at the CSRe.
The results for six $T_{z}=-3/2$ nuclides are in excellent agreement to the previously published data.
However, the mass precision was significantly improved for the ions of interest lying close to the ``good'' isochronous region at $\gamma \sim \gamma_t$.
Furthermore, the method enables a quick and reliable verification of the isochronous ion-optical setting of the ring.

The present method is based on three assumptions:
(1) the revolution periods for each stored ion should, at least approximately, be normally distributed;
(2) At least, two ions should be stored simultaneously in each individual measurement;
and (3) More than three different ion species should randomly occur in various measurements.
These requirements are usually satisfied in all IMS experiments at different storage rings,
therefore this method is suitable in principle for most of isochronous mass measurements,
or even for similar data analyses in other types of experiments.

\section*{Acknowledgments}

This work is supported in part by the 973 Program of China (No. 2013CB834401), the NSFC (Grants No. 11035007, U1232208, and 11205205), the Chinese Academy of Sciences, and BMBF grant in the framework of the Internationale Zusammenarbeit
in Bildung und Forschung (Projekt-Nr. 01DO12012), the External Cooperation Program of the Chinese Academy of Sciences (Grant No. GJHZ1305), and
the Helmholtz-CAS Joint Research Group (Group No. HCJRG-108). Y.A.L is supported by CAS visiting professorship for senior international scientists
(Grant No. 2009J2-23). K.B. and Y.A.L. acknowledge support by the Nuclear Astrophysics Virtual Institute (NAVI) of the Helmholtz Association and thank ESF
for support within the EuroGENESIS program. T.Y. acknowledges support by The Mitsubishi Foundation.

\end{document}